%% file: main.tex
\documentclass[10pt, conference, letterpaper]{IEEEtran}

\pdfoutput=1
\usepackage{listings}

\usepackage{multicol}

\usepackage{blindtext}
\usepackage{amsfonts}
\usepackage{amsmath, amsthm, amssymb}
\usepackage{times}
\usepackage{url}
\usepackage{verbatim}
\usepackage{array}
\usepackage{mathrsfs}
\usepackage{enumitem}
\usepackage{textcomp}

\newcommand{\subparagraph}{}
\usepackage[compact]{titlesec}
\usepackage[ruled, lined, linesnumbered]{algorithm2e}
\SetAlFnt{\small}

\SetCommentSty{mycommfont}

\usepackage{amsfonts}
\usepackage{bbm}
\usepackage{soul}
\usepackage{color}
\usepackage{graphics}
\usepackage{graphicx,epsfig}
\usepackage{lipsum,graphicx,subcaption}
\graphicspath{{figures/}}
\usepackage{caption}
\usepackage{subcaption}

\usepackage{cleveref}
\usepackage{wrapfig}
\usepackage{url}
\usepackage{url}
\usepackage{amsmath,amssymb,epsfig,amsthm,psfrag,epsf, multirow,wrapfig, graphicx}
\usepackage[english]{babel}
\usepackage{epstopdf}
\usepackage{float}
\usepackage{color}
\usepackage[table,xcdraw]{xcolor}
\usepackage{amsmath}
\usepackage{amsfonts}
\usepackage{amssymb}
\usepackage{mathtools}
\DeclarePairedDelimiter{\ceil}{\lceil}{\rceil}
\usepackage{tabularx}
\usepackage{adjustbox}
\usepackage{physics}

\usepackage{amsfonts}
\usepackage{amssymb}
\usepackage{caption}
\usepackage{subcaption}
\usepackage{tabularx}
\usepackage{comment}
\usepackage[mathscr]{euscript}
\usepackage{amsbsy}
\usepackage{stmaryrd}
\usepackage{mathtools}
\usepackage[utf8]{inputenc}
\graphicspath{{figs/}}

\usepackage{booktabs}
\usepackage{graphicx,epsfig}
\usepackage{graphics}
\usepackage{multirow}
\usepackage{rotating}
\usepackage{caption}
\usepackage{subcaption}
\usepackage{amsmath}
\usepackage{amssymb}
\usepackage{makecell}

\DeclareMathOperator*{\argmax}{arg\,max}

\usepackage{mathtools}
\usepackage{comment}
\usepackage{multirow}

\usepackage{footnote}
\usepackage{tablefootnote}
\usepackage{scalerel}
\usepackage{tabularx}
\usepackage{mdframed}
\usepackage{lipsum}
\makeatletter
\newcommand*{\rom}[1]{\expandafter\@slowromancap\romannumeral #1@}
\makeatother
\usepackage{nccmath}

\usepackage{relsize}
\renewcommand\IEEEkeywordsname{Keywords}

\newcommand{\note}[1]{{\color{blue}{#1}}} 
\newcommand{\inred}[1]{{\color{red}{#1}}}

\DeclareMathSymbol{\shortminus}{\mathbin}{AMSa}{"39}

\allowdisplaybreaks
\begin{document}
\def\eg{\mbox{\em e.g.}, }

\title{Proofs and Supplementary Material:\\ Unified Characterization and Precoding for Non-Stationary Channels}%\vspace{-10ex}
% \title{Unified Characterization and Precoding for Non-Stationary Channels}%\vspace{-10ex}
% \title{Precoding Non-Stationary MIMO Channels}%\vspace{-10ex}
%\title{Practical Non-linear Pre-equalization for Non-stationary Channels\\ (VFRP Mid-term Report) \\ }%\vspace{-10ex}

\begin{comment}
% \author{
%     \IEEEauthorblockN{Maqsood Careem, Aveek Dutta and Ngwe Thawdar}
%     \IEEEauthorblockA{Department of Electrical and Computer Engineering\\
%     University at Albany SUNY, Albany, NY 12222 USA
%     \\\{mabdulcareem\}@albany.edu}
%     US Air Force Research Laboratory
% Rome, NY, USA, E-mail: ngwe.thawdar@us.af.mil
%     % \vspace{-3pt}
%     % \vspace{-5ex}
%     % \vspace{-10ex}
% }
\end{comment}

\author{
\begin{tabular}[t]{c@{\extracolsep{8em}}c} 
Zhibin Zou, Maqsood Careem, Aveek Dutta & Ngwe Thawdar \\
Department of Electrical and Computer Engineering & US Air Force Research Laboratory \\ 
University at Albany SUNY, Albany, NY 12222 USA & Rome, NY, USA \\
\{{zzou2, mabdulcareem, adutta\}@albany.edu} & 
ngwe.thawdar@us.af.mil
\end{tabular}
\vspace{-3ex}
}

% \author{
% \IEEEauthorblockN{Zhibin Zou, Maqsood Careem and Aveek Dutta}
%     \IEEEauthorblockA{Department of Electrical and Computer Engineering\\
%     University at Albany SUNY, Albany, NY 12222 USA\\
%     \{zzou2, mabdulcareem, adutta\}@albany.edu}
%     \vspace{-5ex}

% }
    
% \author{}
%\IEEEoverridecommandlockouts
%\IEEEpubid{\makebox[\columnwidth]{978-1-7281-2294-6/19/\$31.00~\copyright2019 IEEE IEEE \hfill} \hspace{\columnsep}\makebox[\columnwidth]{ }}
\maketitle
% \vspace{-3em}
% \vspace{-30pt}
%\IEEEoverridecommandlockouts\IEEEpubid{\makebox[\columnwidth]{978-1-7281-2294-6/19/\$31.00 \textcopyright2019 IEEE \hfill} \hspace{\columnsep}\makebox[\columnwidth]{ }}

% \input{abstract}
% % \vspace{-10pt}
% \input{intro}
% % \input{related_NLP}
% % \input{primer}
% % \newpage
% % \vspace{-10pt}
% \input{background}
% % \newpage
% \input{models}
% % \newpage
% \input{precoding}
% %\input{system}
% %\input{AE_KLT}
% %\input{KLT_precoding}
% %\input{complexity}
% % \newpage
% \input{results}
% % \newpage
% \input{conclusion}
% %\input{intro}
% % \input{problem_statement}
% % \newpage
% %\input{equivalence}
% % \newpage
% %\input{bounds}
% % \input{design}
% % \newpage
% %\input{results}
% %\input{related}
% % \input{channel_estimation}
% %\input{conclusion}
% % \input{new_abstract}
% % \input{new_content.tex}
% % % \input{appendix_new}

% \input{intro.tex}
% \input{model.tex}
% \input{equivalence.tex}
% \input{generalization}
% \input{design.tex}
% \input{data_augmentation.tex}
% \input{implementation.tex}
% \input{results.tex}
% \input{related.tex}
% \input{conclusion.tex}

% \bibliographystyle{IEEEtran}
% \bibliography{references}
%\bibliographystyle{ACM-Reference-Format}

% \newpage

% \input{appendix_NLP.tex}
%\input Appendix1.tex
% \input{appendix_decoding}

%%%% \balancecolsandclearpage
% \clearpage
% \newpage
% \setcounter{page}{1}
% \input{appendix_icc}
%%% \newpage

% \bibliographystyle{IEEEtran}
% \bibliography{references, ref_precoding}

% \input{comment}

% \clearpage
% \newpage
\setcounter{equation}{34}
\input{appendix_NLP}

\bibliographystyle{IEEEtran}
\bibliography{references}

\end{document}

%% file: appendix_NLP.tex
\appendices
\label{appendix:precoding}

% \begin{multicols}{2}
% [
% \begin{center}
% \large{\textbf{Proofs and Supplementary Material:\\ Unified Characterization and Precoding for Non-Stationary Channels}}\\

% \vspace{5pt}
% \normalsize
% Zhibin Zou\textsuperscript{\textbf{*}}, Maqsood Careem\textsuperscript{\textbf{*}}, Aveek Dutta\textsuperscript{\textbf{*}} \& \\
% Ngwe Thawdar\textsuperscript{\textbf{**}} \\

% \vspace{5pt}
% \small
% \textsuperscript{\textbf{*}}Department of Electrical and Computer Engineering,\\ 
% University at Albany SUNY, Albany, NY 12222 USA\\
% \textsuperscript{\textbf{**}}US Air Force Research Laboratory,\\
% Rome, NY, USA
% \end{center}

% \normalsize

\noindent
\textbf{Instructions:}
This document provides the supplementary material including a comprehensive related work, the complete proofs and extended evaluation results that support the manuscript, ``Unified Characterization and Precoding for Non-Stationary Channels", that was accepted for publication at IEEE International Conference on Communications (ICC) 2022.  Equations (1)--(34) refer to the equations from the main manuscript, and the Theorem, Lemma and Corollaries correspond to those from the manuscript. 
% Equations \eqref{eq:H_delay_Doppler}--\eqref{eq:xut} and the references [1]--[23] refer to the equations and references from the main manuscript (``Unified Characterization and Precoding for Non-Stationary Channels" accepted for publication at IEEE ICC 2022), respectively. This document provides the supplementary material including a comprehensive related work, the complete proofs and extended evaluation results to support the main manuscript.  

% \noindent
% % ]
% \end{multicols}

\input{related_NLP}

\section{Proofs on Unified Characterization}
\label{app:characterization}

\subsection{Proof of Lemma 1: Generalized Mercer's Theorem}
\label{App:gmt}

\begin{proof}
Consider a 2-D process $K(t,t') \in L^2(Y \times X)$, where $Y(t)$ and $X(t')$ are square-integrable zero-mean random processes with covariance function $K_{Y}$ and $K_{X}$, respecly. 
The projection of $K(t, t')$ onto $X(t')$ is obtained as in \eqref{eq:projection},
\begin{align}
\label{eq:projection}
    & C(t) = \int K(t, t') X(t') ~dt'
\end{align}

Using \textit{Karhunen–Loève Transform} (KLT), $X(t')$ and $C(t)$ are both decomposed as in \eqref{eq:X_t} and \eqref{eq:C_t}, 
\begin{align}
    &X(t') = \sum_{i = 1}^{\infty} x_{i} \phi_{i}(t') \label{eq:X_t}\\
    &C(t) = \sum_{j = 1}^{\infty} c_{j} \psi_{j}(t) \label{eq:C_t}
\end{align}
where $x_i$ and $c_j$ are both random variables with $\mathbb{E}\{x_i x_{i'}\} {=} \lambda_{x_i} \delta_{ii'}$ and $\mathbb{E}\{c_j c_{j'}\} {=} \lambda_{c_j} \delta_{jj'}$.  $\{\lambda_{x_i}\}$, $\{\lambda_{x_j}\}$ $\{\phi_i(t')\}$ and $\{\psi_j(t)\}$ are eigenvalues and eigenfuncions, respectively.
% of $T_{K_{X}}$ and $T_{K_{C}}$, respectively. 
% 
% 
Let us denote $n{=}i{=}j$ and $\sigma_n {=} \frac{c_n}{x_n}$, and assume that $K(t,t')$ can be expressed as in \eqref{eq:thm_K_t},
\begin{align}
\label{eq:thm_K_t}
    K(t,t') = \sum_n^\infty \sigma_n \psi_{n}(t) \phi_{n}(t')
\end{align}
We show that \eqref{eq:thm_K_t} is a correct representation of $K(t,t')$ by proving \eqref{eq:projection} holds under this definition. 
We observe that by substituting \eqref{eq:X_t} and \eqref{eq:thm_K_t} into the right hand side of \eqref{eq:projection} we have that,
\begin{align}
    & \int K(t, t') X(t') ~dt' \nonumber \\
    & = \int \sum_n^\infty \sigma_n \psi_{n}(t) \phi_{n}(t') \sum_{n}^{\infty} x_{n} \phi_{n}(t') ~dt' \nonumber \\
    & = \int \sum_n^\infty \sigma_n x_n \psi_n(t) |\phi_n(t')|^2 \nonumber\\
    & + \sum_{n'\neq n}^ \infty \sigma_{n} x_{n'} \psi_{n}(t) \phi_{n}(t') \phi_{n'}^*(t') ~d t' \nonumber \\
    & = \sum_n^\infty c_n \psi_n(t) = C(t)
\end{align}
which is equal to the left hand side of \eqref{eq:projection}. 
Therefore, \eqref{eq:thm_K_t} is a correct representation of $K(t,t')$.

\begin{comment}
Substitute the left side of \eqref{eq:thm1_1} by \eqref{eq:thm1_2}, we have 
\begin{equation}
    \sum_{j = 1}^{\infty} \mu_{j} \psi_{j}(t) = \sum_{i = 1}^{\infty} z_i c_i(t)
\end{equation}

Notice $i$ and $j$ have the same range. Denote $n = i = j$ and $\sigma_n = \mu_n / z_n$, we have
\begin{align}
    &\sigma_n \psi_{n}(t) =  c_n(t) \nonumber \\ 
    & = \int K(t, t') \phi_{n}(t') ~dt'
\end{align}

Thus $K(t, t')$ can be decomposed
\begin{align}
    K(t, t') = \sum_{n=1}^{\infty} \sigma_n \psi_n(t) \phi_n(t') 
\end{align}
where $E\{\sigma_n \sigma_n'\} = \lambda_n \delta_{nn'}$. $\lambda_n$ is eigenvalue. $\psi_n(t)$ and $\phi_n(t')$ are both eigenfunctions.
\end{comment}

\begin{comment}
Consider a P-dimensional process $K \in L^P(X_1 \times X_2\times \cdots \times X_P)$.

For a square-integrable zero-mean random process process $X_i, i\in \{1, 2, \cdots, P\}$, with covariance function $K_{X_i}$ Karhunen–Loève theorem gives 

\begin{equation}
    X_i = \sum_{n}^{\infty} \sigma_{n}^{(i)} \phi_{n}^{(i)}(t_i)
\end{equation}
where $E\{\sigma_n^{(i)} \sigma_n'^{(i)}\} = \lambda_n^{(i)} \delta_{nn'}$. $\lambda_n^{(i)}$ and $\phi_n^{(i)}(t_i)$ are eigenvalue and eigenfuncion of $T_{K_{X_i}}$, respectively. 
\end{comment}
\end{proof}

\subsection{Proof of Theorem 1: High Order Generalized Mercer's Theorem (HOGMT}
\label{app:hogmt}

\begin{proof}
Given a 2-D process $X(\gamma_1, \gamma_2)$, the eigen-decomposition using Lemma 1 is given by,
\begin{equation}
\label{eq:thm1_1}
    X(\gamma_1, \gamma_2) = \sum_{n}^{\infty} x_{n} e_n(\gamma_1) s_n(\gamma_2)
\end{equation}

Letting $\psi_n(\gamma_1,\gamma_2) {=} e_n(\gamma_1) s_n(\gamma_2)$, and substituting it in \eqref{eq:thm1_1} we have that,

\begin{equation}
\label{eq:2d_klt}
    X(\gamma_1, \gamma_2) = \sum_{n}^{\infty} x_{n} \phi_n(\gamma_1,\gamma_2)
\end{equation}
where $\phi_n(\gamma_1,\gamma_2)$ are 2-D eigenfunctions with the property \eqref{eq:prop1}.
\begin{equation}
\label{eq:prop1}
\iint \phi_n(\gamma_1,\gamma_2) \phi_{n'}(\gamma_1,\gamma_2) ~d\gamma_1 ~d\gamma_2 = \delta_{nn'} 
\end{equation}

We observe that \eqref{eq:2d_klt} is the 2-D form of KLT. With iterations of the above steps, we obtain \textit{Higher-Order KLT} for $X(\gamma_1,\cdots,\gamma_Q)$ and $C(\zeta_1,\cdots,\zeta_P)$ as given by,
\begin{align}
   & X(\gamma_1,\cdots,\gamma_Q) = \sum_{n}^{\infty} x_{n} \phi_n(\gamma_1,\cdots,\gamma_Q) \\
   & C(\zeta_1,\cdots,\zeta_P) = \sum_{n}^{\infty} c_{n} \psi_n(\zeta_1,\cdots,\zeta_P)
\end{align}
where $C(\zeta_1,\cdots,\zeta_P)$ is the projection of $X(\gamma_1,\cdots,\gamma_Q)$ onto $K(\zeta_1,\cdots,\zeta_P; \gamma_1,\cdots, \gamma_Q)$.

Then following similar steps as in Appendix~\ref{App:gmt} we get \eqref{eq:col}. 
\begin{align}
\label{eq:col}
& K(\zeta_1,\cdots,\zeta_P; \gamma_1,\cdots, \gamma_Q) \nonumber \\
& = \sum_{n}^ \infty \sigma_n \psi_n(\zeta_1,\cdots,\zeta_P) \phi_n(\gamma_1,\cdots, \gamma_Q)
\end{align}
\end{proof}

\section{Proofs on Eigenfunction based Precoding}
\label{app:precoding}

\subsection{Proof of Lemma 2}
\label{app:lem1}

\begin{proof}
Using 2-D KLT as in (13), $x(u,t)$ is expressed as,
\begin{equation}
    x(u,t) = \sum_{n}^ \infty x_n \phi_n(u,t)
\end{equation}
where $x_n$ is a random variable with $E\{x_n x_{n'}\}{=} \lambda_n \sigma_{nn'} $ and $\phi_n(u,t)$ is a 2-D eigenfunction. 

\begin{comment}
\begin{align}
    & E\{|x(u,t)|^2\} = E\{|\sum_{n}^ \infty x_n \phi_n(u,t)|^2\} \nonumber\\
    & = \sum_{n}^ \infty E\{|x_n|^2\} E\{|\phi_n(u,t)|^2\} \nonumber \\
    & = \sum_{n}^ \infty \frac{\lambda_n}{T}\iint_{T}|\phi_n(u,t)|^2 ~du ~dt  \nonumber \\
    & = \sum_{n}^ \infty \frac{\lambda_n}{T} 
\end{align}
where $T$ is the time interval. Then minimizing $E\{|x(t)|^2\}$ converts to minimizing $\sum_j^ \infty \lambda_j$.
\end{comment}

Then the projection of $k_H(u,t;u',t')$ onto $\phi_n(u',t')$ is denoted by $ c_n(u,t)$ and is given by,
\begin{equation}
    c_n(u,t) =  \iint k_H(u,t;u',t') \phi_n(u',t') ~du' ~dt'
\end{equation}

Using the above, (28) is expressed as,
\begin{align}
\label{eq:obj_trans}
     & ||s(u,t) - Hx(u,t)||^2 = ||s(u,t) - \sum_n^ \infty x_n c_n(u,t)||^2
\end{align}

Let $\epsilon (x) {=} ||s(u,t) - \sum_n^ \infty x_n \phi_n(u,t)||^2$. Then its expansion is given by,
\begin{align}
\label{eq:ep}
     & \epsilon (x) = \langle s(u,t),s(u,t) \rangle - 2\sum_n ^ \infty  x_n \langle c_n(u,t),s(u,t) \rangle \\
     & + \sum_n^ \infty x_n^2 \langle c_n(u,t), c_n(u,t) \rangle \nonumber + \sum_n^ \infty \sum_{n' \neq n}^ \infty x_n x_{n'}  \langle c_n(u,t), c_{n'}(u,t) \rangle
\end{align}

Then the solution to achieve minimal $\epsilon(x)$ is obtained by solving for $\pdv{\epsilon(x)}{x_n} = 0$ as in \eqref{eq:solution}.
\begin{align}
\label{eq:solution}
    x_n^{opt} & {=}  \frac{\langle s(u,t), c_n(u,t) \rangle + \sum_{n'\neq n}^ \infty x_{n'} \langle c_{n'}(u,t), c_n(u,t) \rangle }{\langle c_n(u,t), c_n(u,t) \rangle}
\end{align}
where $\langle a(u,t), b(u,t) \rangle {=} \iint a(u,t) b^*(u,t) ~du ~dt$ denotes the inner product. 
Let $\langle c_{n'}(u,t), c_n(u,t) \rangle = 0$, i.e., the projections $\{ c_n(u,t)\}_n$ are orthogonal basis. Then we have a closed form expression for $x^{opt}$ as in \eqref{eq:x_opt}.
\begin{align}
\label{eq:x_opt}
    x_n^{opt} & {=}  \frac{\langle s(u,t), c_n(u,t) \rangle}{\langle c_n(u,t), c_n(u,t) \rangle}
\end{align}

Substitute \eqref{eq:x_opt} in \eqref{eq:ep}, it is straightforward to show that $\epsilon(x){=} 0$.
\end{proof}

\subsection{Proof of Theorem 2: Eigenfunction Precoding}
\label{app:thm_2}
\begin{proof}
The 4-D kernel $k_H(u,t;u',t')$ is decomposed into two separate sets of eigenfunction $\{\phi_n(u',t')\}$ and $\{\psi_n(u, t) \}$ using Theorem 1 as in (30). By transmitting the conjugate of the eigenfunctions, $\phi_n(u',t')$ through the channel $H$, we have that,  
\begin{align}
   & H \phi_n^*(u',t') = \iint k_H(u,t;u',t') \phi_n^*(u',t') ~du' ~d t' \nonumber \\ 
   & {=} \iint \sum_{n}^ \infty \{\sigma_n \psi_n(u,t) \phi_n(u',t')\} \phi_n^*(u',t') ~d t' ~d f' \nonumber \\ 
   & {=} \iint \sigma_n \psi_n(u,t) |\phi_n(u',t')|^2 \nonumber\\
   & + \sum_{n'\neq n}^ \infty \sigma_{n'} \psi_{n'}(u,t) \phi_{n'}(u',t')\ \phi_n^*(u',t') ~du' ~d t' \nonumber \\
   & {=} \sigma_n \psi_n(u,t)
\end{align}
where $\psi_n(u,t)$ is also a 2-D eigenfunction with the orthogonal property as in (31). 

From Lemma 2, if the set of projections, $\{c_n(u,t)\}$ is the set of eigenfunctions, $\{\psi_n(u,t)\}$, which has the above orthogonal property, we achieve the optimal solution as in \eqref{eq:x_opt}. Therefore, let $x(u,t)$ be the linear combination of $\{\phi_n^*(u,t)\}$ with coefficients $\{x_n\}$ as in \eqref{eq:construct},

\begin{equation}
\label{eq:construct}
    x(u,t) = \sum_n^ \infty x_n \phi_n^*(u,t) 
\end{equation}

Then \eqref{eq:obj_trans} is rewritten as in \eqref{eq:obj_trans2},
\begin{align}
\label{eq:obj_trans2}
     & ||s(u,t) - Hx(u,t)||^2 = ||s(u,t) - \sum_n^ \infty x_n \sigma_n \psi(u,t)||^2
\end{align}
   
Therefore, optimal $x_n$ in \eqref{eq:x_opt} is obtained as in \eqref{eq:opt},

\begin{equation}
\label{eq:opt}
    x_n^{opt} = \frac{\langle s(u,t), \psi_n(u,t) \rangle}{\sigma_n} 
\end{equation}

Substituting \eqref{eq:opt} in \eqref{eq:construct}, the transmit signal is given by \eqref{eq:x_opt2},
\begin{equation}
\label{eq:x_opt2}
    x(u,t) = \sum_n^ \infty \frac{\langle s(u,t), \psi_n(u,t) \rangle}{\sigma_n} \phi_n^*(u,t). 
\end{equation}
\end{proof}

\subsection{Proof of Corollary 1}
\label{app:EP_space}
\begin{proof}
First we substitute the 4-D kernel $k_H(u,t;u',t')$ with the 2-D kernel $k_H(u,u')$ in Theorem 2 which is then decomposed by the 2-D HOGMT. Then following similar steps as in Appendix~\ref{app:thm_2} it is straightforward to show (34).
\end{proof}

\section{Results on Interference}
\label{App:results_interference}
\begin{figure}[h]
  \centering
  \includegraphics[width=1\linewidth]{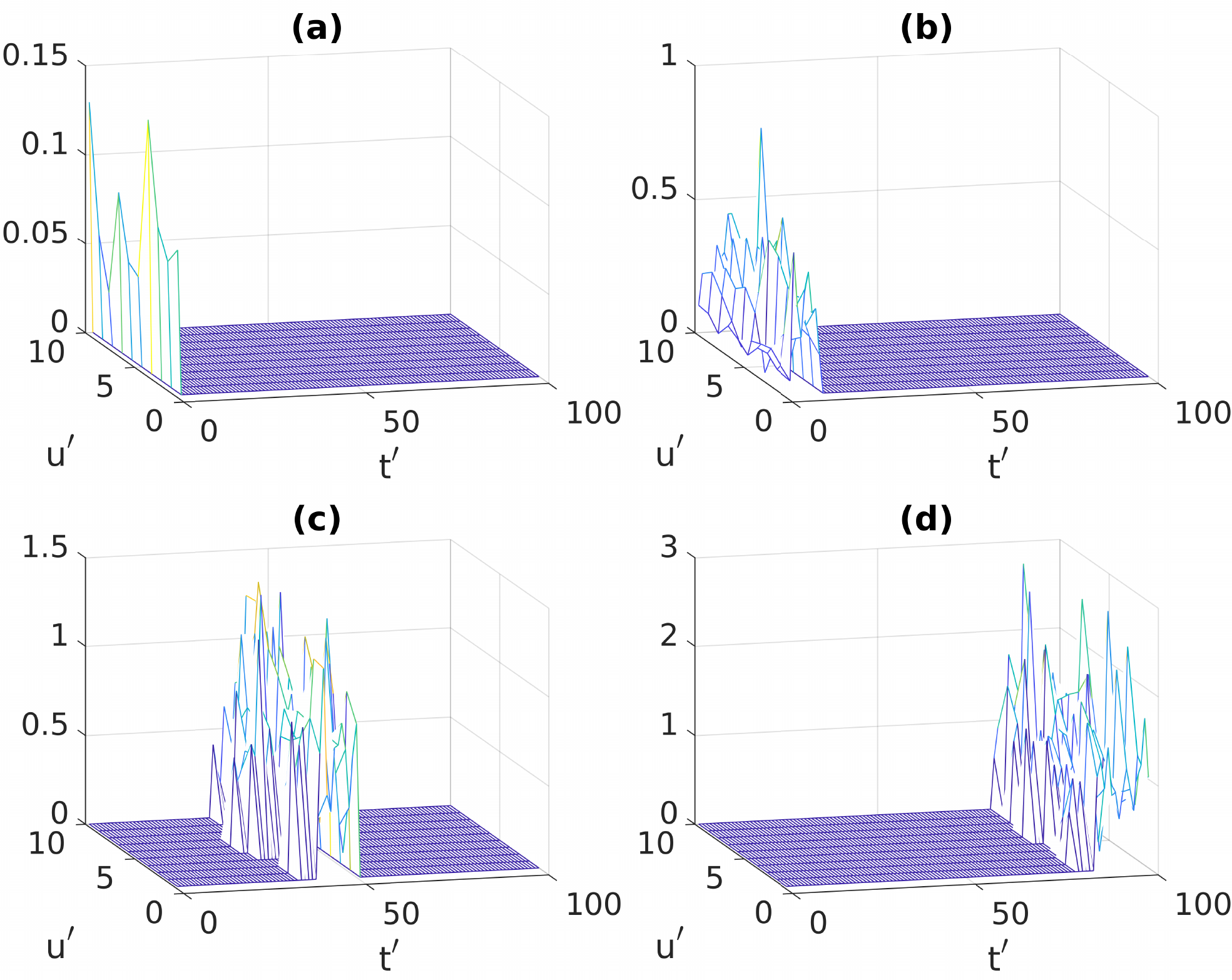}
  \caption{Kernel $k_H(u,t;u',t')$ for $u {=} 1$ at a) $t {=} 1$, b) $t {=} 10 $, c) $t {=} 50$ and d) $t {=} 100$.}
%   \note{Include z label and same scale.}}
  \label{fig:hst_1_10_50_100}
\label{fig:hst_1_10_50_100}
\end{figure}
Figure~\ref{fig:hst_1_10_50_100} shows the channel response for user $u {=} 1$ at $t{=}1$, $t{=}10$, $t{=}50$ and $t{=}100$, where at each instance, the response for user $u {=} 1$ is not only affected by its own delay and other users' spatial interference, but also affected by other users' delayed symbols. 
% We call it space-time joint interference.  
This is the cause of joint space-time interference which necessitates joint precoding in the 2-dimensional space using eigenfunctions that are jointly orthogonal.

%% file: related_NLP.tex
\section{Related work}
\label{App:related_NLP}

% \inred{Maqsood: Trim only to material relevant to this work.}

We categorize the related work into three categories:

\noindent
\textbf{Characterization of Non-Stationary Channels:}
% \note{Zhibin: Include relevant references...}
% Although many literature attempt to characterize the channel they are often impractical as they rely on several statistics, by we do something different... Add citations and more context.
Wireless channel characterization in the literature typically require several local and global (in space-time dimensions) higher order statistics to characterize or model non-stationary channels, due to their time-varying statistics. 
These statistics cannot completely characterize the non-stationary channel, however are useful in reporting certain properties that are required for the application of interest such as channel modeling, assessing the degree of stationarity etc.
Contrarily, we leverage the 2-dimensional eigenfunctions that are decomposed from the most generic representation of any wireless channel as a spatio-temporal channel kernel.
These spatio-temporal eigenfunctions can be used to extract any higher order statistics of the channel as demonstrated in Section \rom{3}, and hence serves as a complete characterization of the channel.
Furthermore, since this characterization can also generalize to stationary channels, it is a unified characterization for any wireless channel.
Beyond characterizing the channel, these eigenfunctions are the core of the precoding algorithm.
% Therefore, this serves as a complete characterization of any non-stationary 
% Decomposition is essential for precoding but also allows us to completely characterize the wireless channel by explaining its statistics, without having to calculate all the statistics as in the literature. 
% Additionally, the decomposition that is most useful to ensure interference-free communication is one that preserves orthogonality ... Why is 4-D to 2-D decomposition important?

\noindent
\textbf{Precoding Non-Stationary Channels:}
Although precoding non-stationary channels is unprecedented in the literature \cite{AliNS0219}, we list the most related literature for completeness. 
The challenge in precoding non-stationary channels is the lack of accurate models of the channel and the (occasional) CSI feedback does not fully characterize the non-stationarities in its statistics. This leads to suboptimal performance using state-of-the-art precoding techniques like Dirty Paper Coding which assume that complete and accurate knowledge of the channel is available, while the CSI is often outdated in non-stationary channels. 
While recent literature present  attempt to deal with imperfect CSI by modeling the error in the CSI \cite{HatakawaNLP2012, HasegawaTHP2018, GuoTHP2020, DietrichTHP2007, CastanheiraPGS2013, WangTHP2012, MazroueiTDVP2016, Jacobsson1DAC2017}, they are limited by the assumption the channel or error statistics are stationary or WSSUS at best.
Another class of literature, attempt to deal with the impact of outdated CSI~\cite{AndersonLP2008,Zeng2012LP} in time-varying channels by quantifying this loss or relying statistical CSI. These methods are not directly suitable for non-stationary channels, as the time dependence of the statistics may render the CSI (or its statistics) stale, consequently resulting in precoding error. 

\noindent
\textbf{Space-Temporal Precoding:}
While, precoding has garnered significant research, spatio-temporal interference is typically treated as two separate problems, where spatial precoding at the transmitter aims to cancel inter-user and inter-antenna interference, while equalization at the receiver mitigates inter-carrier and inter-symbol interference.
% \ingreen{Alternately, \cite{hadani2018OTFS} proposes an orthogonal time frequency OTFS modulation scheme that make symbols experience minimal cross-interference by signal processing at the transmitter and receiver. However, equalizer is still needed to cancel 2-D intersymbol interference especially for non-staitonary channels.}
Alternately, \cite{hadani2018OTFS} proposes to modulate the symbols such that it reduces the cross-symbol interference in the delay-Doppler domain, but requires equalization at the receiver to completely cancel such interference in practical systems. %by introducing additional signal processing at the transmitter and receiver. 
Moreover, this approach cannot completely minimize the joint spatio-temporal interference that occurs in non-stationary channels since their statistics depend on the time-frequency domain in addition to the delay-Dopper domain (explained in Section \rom{2}).
% , and it requires equalization at the receiver to completely cancel the cross-symbol interference in the delay-Doppler domain for practical systems \cite{hadani2018OTFS}.
While spatio-temporal block coding techniques are studied in the literature \cite{Cho2010MIMObook} they add redundancy and hence incur a communication overhead to mitigate interference, which we avoid by precoding. 
These techniques are capable of independently canceling the interference in each domain, however are incapable of mitigating interference that occurs in the joint spatio-temporal domain in non-stationary channels. 
We design a joint spatio-temporal precoding that leverages the extracted 2-D eigenfunctions from non-stationary channels to mitigate interference that occurs on the joint space-time dimensions, which to the best of our knowledge is unprecedented in the literature.